Special Issue (SI):  MARC X
LOG NUMBER OF PAPER: 136
TITLE OF PAPER: HIGH QUALITY ACTINIDE TARGETS
AUTHOR(S): W. LOVELAND
POSTAL ADDRESS OF EACH AUTHOR: B123 RADIATION CENTER, OREGON STATE UNIVERSITY, CORVALLIS, OR 97331 USA
CORRESPONDING AUTHOR'S E-MAIL ADDRESS, TELEPHONE AND FAX NUMBERS:

lovelanw@onid.orst.edu
541-737-7078
Fax  541-7737-0480

# HIGH QUALITY ACTINIDE TARGETS

W. Loveland

Dept. of Chemistry

Oregon State University

Corvallis, OR 97331 USA

## ABSTRACT

We prepare high quality actinide targets for studies of neutron-induced and charged-particle-induced fission. I report on our efforts to measure fragment energy loss in the target backings and to diagnose the "crud" problem frequently found in $^{248}$Cm and $^{252}$Cf sources and targets. I discuss the preparation of multi-isotopic targets for the FissionTPC and our efforts to measure the pointing resolution of this device. The issues of target uniformity, chemical composition and radiation stability of the targets are discussed along with problems of high/low specific activity regions in a single target.

## KEYWORDS

## INTRODUCTION

Our research program involves studies of neutron induced [1,2] and charged particle induced fission [3,4]. While the goals of these research programs are quite different, they all involve carefully prepared, high quality targets of the actinide elements. All studies involve thin (50-300 µg/cm$^2$), uniform deposits of actinide compounds on thin (< 100 µg/cm$^2$) backing materials such as C, Al, Ti or polymeric materials. Target activities range from Bq ($^{238}$U, $^{232}$Th) to MBq ($^{239}$Pu, $^{252}$Cf). The "comfort level" of the people handling the targets, i.e., their estimation of the probability of target breakage and its consequences, largely set the thicknesses of the target backings. Consequently, thicknesses of target backing materials are ~ 50 – 100 µg/cm$^2$.

A number of the targets we produce are intended for use with the Fission Time Projection Chamber [1], a unique tracking device for studying fission. The current thrust of the FissionTPC scientific program is to measure neutron induced fission cross sections for $^{235,238}$U and $^{239}$Pu in the neutron energy range from 1 to 100 MeV with a precision and accuracy of < 1%. One of the keys to achieving this daunting goal is to have high quality actinide targets for the device.

As part of the renaissance of interest in fission, we are also engaged in measurements of the total kinetic energy release in fission, its variance and the dependence of these quantities upon the excitation energy of the fissioning system, E*.

Our studies of charged particle induced fission [3,4] are part of our general program to understand the synthesis of the heaviest elements, where fission is a stability-limiting factor.

**GENERAL CONSIDERATIONS**

While our research projects are quite different, we have realized certain general principles guide high quality actinide target preparation. For charged particle irradiations, we prefer Ti backings as the coefficient of thermal expansion of Ti is similar to that of $UF_4$ [5]. Carbon is preferred as a target backing material for (n, f) studies because it is easily obtainable as a pure thin material that is easy to handle and has good electrical and thermal conductivity properties. We eschew the use of any anti-transfer coatings of the actinide deposits as adding another source of energy loss for fragments leaving the target or in the worst case (Au), serving as a source of fission when bombarded with energetic heavy ions. We prefer the use of vacuum volatilization for making actinide deposits on backing materials due to the superior uniformity of these deposits and the well-characterized form of the deposits ($AnF_4$). However, molecular plating has yields approaching 90-100% while the yields from vacuum volatilization are much less, making molecular plating the method of choice for situations where the actinide material is scarce.

**ISOTOPIC PURITY AND UNIFORMITY**

We are very fortunate that some years ago, we were able to obtain significant quantities of highly enriched $^{235}U$, $^{238}U$ and $^{239}Pu$ (99.9+% enrichment). These materials were obtained from private stocks at U.S. national laboratories. These isotopes were furnished as oxides and we have had to prepare other chemical compounds by synthesis from these stock materials. For example, we have prepared $UF_4$ from $UO_2$ by aqueous synthesis [6,7]. This synthesis leaves us with a hydrated form of $UF_4$ that we heat under vacuum to remove the excess water.

The uniformity of the actinide deposits is determined by autoradiography. For targets prepared by vacuum volatilization, we observe less than a 1.5% variation in the target areal density over the typical charged particle beam spot diameter of 1 cm. For targets prepared by molecular plating, the uniformity is much worse, usually being 10 ± 5% over the same beam spot area.

For targets prepared for use in the FissionTPC, one must remember that the TPC self-autoradiographs all targets using the alpha-activity of the actinide deposits. One can map out and correct for any local non-uniformities in target areal density.

**RADIATION STABILITY OF TARGETS**

In studies of nuclear reactions involving heavy ion collisions with actinide targets, the irradiation conditions pose the possibility of significant damage to the targets. Beam intensities of particle microamperes are commonly used, resulting in ion doses to the targets of $10^{18}$-$10^{19}$. Discoloration of the targets is usually observed with possible loss of actinide material. We conducted a detailed examination of the changes in the surface composition and morphology of $^{238}UF_4$ irradiated with 5 x $10^{18}$ $^{37}Cl$ ions (at an energy of 195 MeV) [8]. We observed changes in the physical structure of the actinide deposits that indicated the actinide material became molten during the irradiation and underwent material flow. (This work was done using AFM imaging and electron microscopy). There was some indication of sputtering of material from the target, with the sputtered material having a composition of $U_2F_x$ with x=7-8. Clearly, in these measurements, one must carefully monitor the targets for possible changes in composition during the experiment.

When one considers neutron-induced fission such as studies using the FissionTPC or measurement of TKE release, the situation is not clear. These experiments are typically carried out at the Weapons Neutron Research Facility (WNR) at the Los Alamos Neutron Science Center (LANSCE) at the Los Alamos National Laboratory. "White spectrum" neutron beams are generated from an unmoderated tungsten spallation source using 800 MeV protons. The neutron energies range from 0.1 to 800 MeV.

To test possible damage to actinide materials under these conditions, we prepared two $^{238}U$ targets with thicknesses of 160 and 280 µg/cm$^2$, coated with 4.5 µg/cm$^2$ Ti. These targets were irradiated with an epi-thermal neutron flux, receiving a dose that was 10x the fast neutron dose expected in one week of irradiation at LANSCE.

The foils changed color during the irradiation and became somewhat more brittle. "No" material loss was observed, i.e., the material loss was < 0.5 %.

**CHEMICAL COMPOSITION OF TARGETS**

In many cases the chemical composition of actinide targets is not well known. Whether that represents a difficulty in using these targets is not clear. For example, with actinide targets prepared by vacuum volatilization, one generally expects (and observes [8]) that the actinide deposit has the formula AnF4 and the backing material is well characterized. The chemical composition of these targets seems to be adequately described. The same cannot be said of targets prepared by molecular plating [9]. The chemical composition of these targets is complex and not well known. In the case of U deposits prepared by molecular plating, Sadi et al. [9] suggest the presence of U-O double bonds, indicating the presence of U (VI). A proposed structure of the molecular plating deposit is shown in Figure 1.

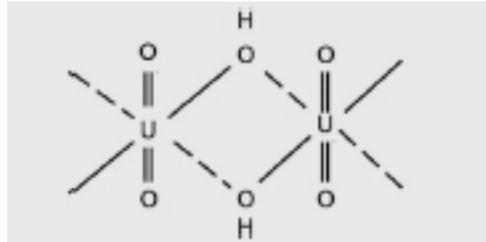

Figure 1. Proposed structure of molecular plating deposit in U targets [9]

Why does this structure matter? It matters if one tries to model or correct for the energy loss of fission fragments emerging from the target. The principal loss of energy is in the fission fragment-light element interactions not in the fragment-heavy element interactions. This problem, along with the 'crud" problem described below are serious impediments for correctly describing the kinetic energy of the emerging fission fragments.

**POINTING RESOLUTION OF THE TPC**

The FissionTPC is a $4\pi$-tracking device. It measures the trajectory of all particles emerging from the target during an experiment. It is crucial that the tracking resolution of the FissionTPC be adequate for this purpose. Design estimates for the Fission TPB predicted that the nominal pointing resolution of the TPC tracking algorithm would be 379 µm with a nominal track angle resolution of 37 mrad. [10]. After construction, one had to determine these quantities. The target makers (us) had to produce extremely complex $^{252}$Cf sources (shown in Figure 2) to test these predictions. These sources were produced by molecular plating $^{252}$Cf deposits using non-conductive masks made by high precision laser machining.

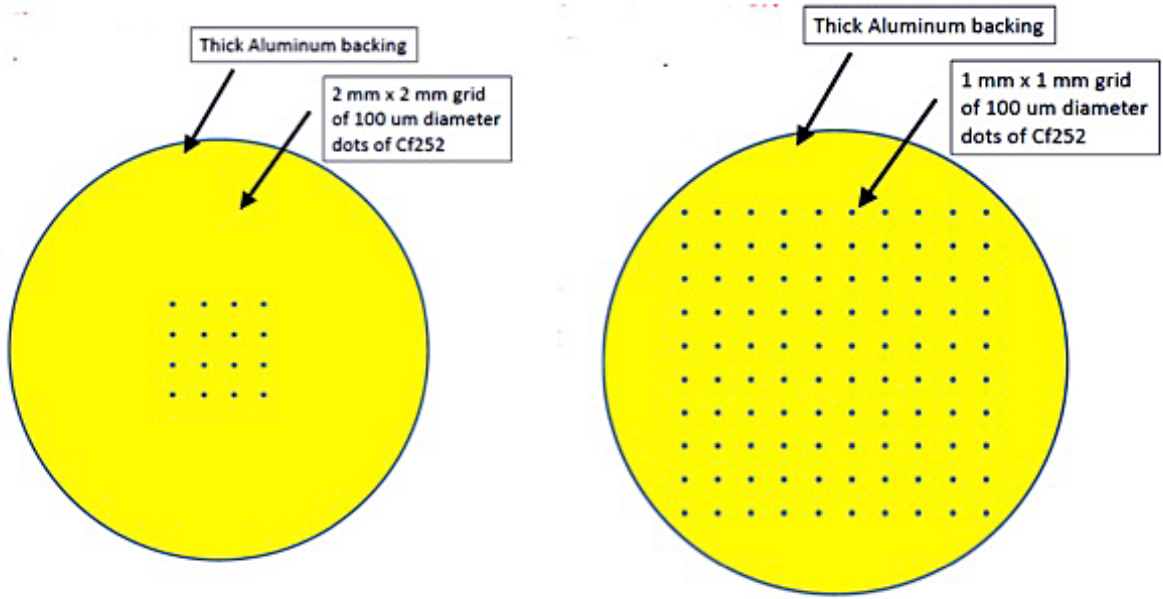

Figure 2. Position resolution $^{252}$Cf targets

**MULTI-ISOTOPIC TARGETS**

One of the real advantages of the FissionTPC is the ability to track each fission event to its point of origin in the target and thus allow the use of multi-isotopic targets. Thus one can simultaneously measure several fission cross sections under identical conditions, removing one of the principal sources of uncertainty in such measurements. The concept is illustrated in Figure 3 where in (a) one shows a target with $^{232}$Th and $^{238}$U wedges or (b) wedges of $^{239}$Pu (left), $^{235}$U and $^{238}$U placed together. One can also have wedges of differing thickness on a target, allowing one to measure the effect of this variable easily.

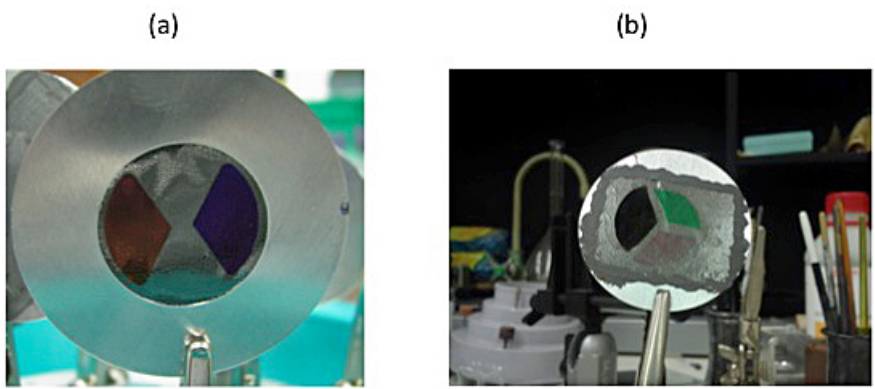

Figure 3. Examples of multi-isotopic target

Preparation of these targets is challenging in that spacing between the wedges must be carefully made, to allow separation between the wedges for tracking but also to allow the wedges to be exposed to similar flux regions in the beam spot. $^{239}$Pu wedges typically have ~ MBq activities while the $^{235,238}$U wedges have Bq levels of activity. One must be vigilant about material creep between the wedges due to self-transfer. Because the TPC tracks all events, ie., fission and α-emission, data collection rates can approach TB/day.

Typically the first wedge to be deposited on a target is the $^{239}$Pu wedge, using molecular plating. This is the most challenging deposition and involves depositing the $^{239}$Pu on a C foil fixed to a glass slide. After this deposition, the C foil is floated off the glass slide and picked up on a target frame. The remaining wedges are deposited by vacuum volatilization (using masking) on the C foil.

**FRAGMENT ENERGY LOSS IN THE TARGET AND TARGET BACKINGS**

The largest single correction when measuring the energy of fission fragments produced in neutron or charged particle induced fission is the correction for fragment energy loss in the target and target backing. Some experimenters make this correction using the widely circulated SRIM program [11]. However this is not a good choice. For example, Knyazheva et al. [12] have shown that up to 30% errors exist in SRIM predictions of the stopping power of heavy fragments in light stoppers. To correct for this effect, we shine a beam of $^{252}$Cf fission fragments upon our targets and measure the energy loss of the average light and heavy fragments, which is then used to iteratively correct the SRIM estimates for stopping power.

**THE "CRUD" PROBLEM**

In molecular plating or electro deposition of actinide materials, one sometimes finds the sample to contain a low mass impurity, simply referred to as "crud". Some have speculated that this impurity is $SiO_2$, which is leached from glass containers of actinide solutions, where the pH is low. The effect of this "crud" is to severely degrade the fission fragment energy spectra. (Fig. 4). (We diagnose the existence of this problem by shining a beam of $^{252}$Cf fission fragments through every actinide deposit we produce, looking for unusual fragment energy losses, prior to using the targets in a real experiment.) This step, while time consuming, allows us to avoid compromised experiments or beam time loss due to poor target performance.

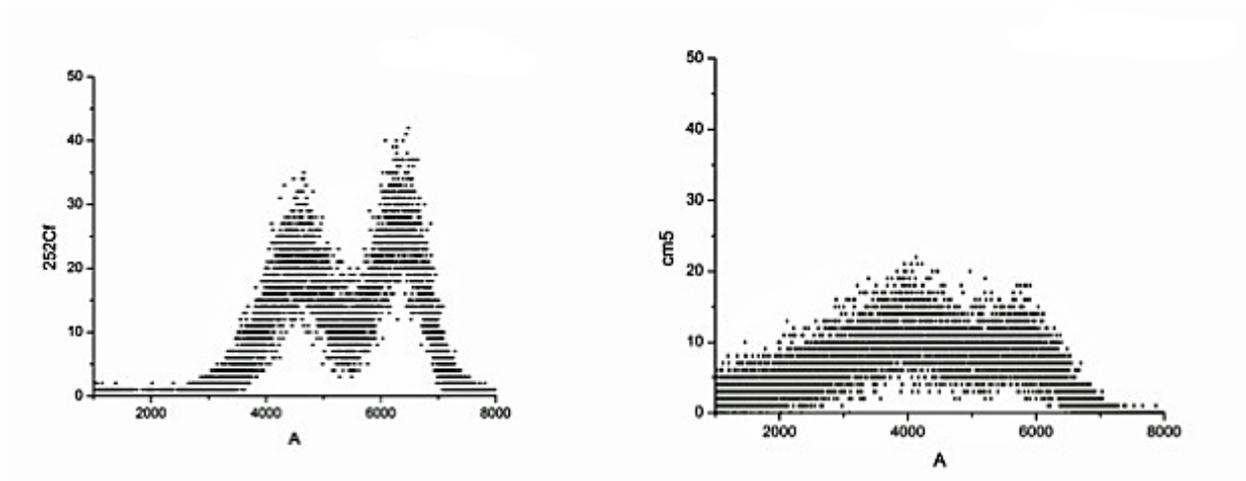

Figure 4. The effect of "crud" on fission spectra. The left hand panel shows a normal $^{252}$Cf spectrum while the right hand panel shows the effect of "crud" on the spectrum.

**CONCLUSIONS**

We conclude that:

- Making high quality actinide targets involves a lot more effort than just making the targets
- Target makers need to participate in the evaluation of the targets and to aid in the interpretation of the results of experiments.
- The FissionTPC, while offering new opportunities in measuring fission cross sections with high accuracy and precision, requires special efforts to create multi-isotopic targets and to demonstrate the position and angular resolution of the device.
- The general problem of evaluating fission fragment energy losses in target materials is best treated by measurements of target properties to guide the interpretation of experimental results.

**ACKNOWLEDGEMENTS**

This work was supported in part by the U.S. Dept. of Energy, Office of Science, Office of Nuclear Physics under Grant DE-FG06-97ER-41026 and the Lawrence Livermore National Laboratory (funding for the FissionTPC project).

**REFERENCES**

1. Heffner M, et al. (2014) Nucl Instru Meth Phys Res A 759:50-64
2. Yanez R, Yao L, King J, Loveland W, Tovesson F, Fotiades N (2014) Phys Rev C 89: 051604

3.  Yanez R, Loveland W, Yao L, Barrett J S, Zhu S, Back B B, Khoo T L, Alcorta M, Albers M (2014) Phys Rev Lett 112: 152702
4.  Yanez R, Loveland W, Barrett J S, Yao L, Back B B, Zhu S, Khoo T L (2013) Phys Rev C 88: 014606
5.  Thermal expansion of UF4
6.  Allen R G, Petrow H G, Magno P J (1958) Ind Eng Chem 50:1748-1749
7.  Allen R G, Petrow H G (1962) U. S. Patent Office 3,023,078
8.  Watson P R, Loveland W, Zielinski P M, Gregorich K E, Nitsche H (2004) Nucl Instru Meth Phys Res B 226:543-548
9.  Sadi S, Paulenova A, Watson P R, Loveland W (2011) Nucl Instru Meth Phys Res A 655: 80-84
10. Heffner M D, Barnes P D, Klay J L (2005) Innovative Fission Measurements with a Time Projection Chamber, UCRL-TR-217600
11. Ziegler J F, Biersack J P, Marwick D J, Cuomo G A, Parker W A, Harrison S A (2003) Available from http://www.srim.org/
12. Kynazheva G N, Khlebnikov S V, Kozulin E M, Kuzmina T E, Lyapin V G, Mutterer M, Perkowski J, Traska W H (2006) Nucl Instru Meth Phys Res B 248:7-15